# Entropy Basis for the Thermodynamic Scaling of the Dynamics of OTP

C M Roland[1] and R Casalini[1,2]

[1] Naval Research Laboratory, Chemistry Division, Code 6120, Washington DC 20375-5342.
[2] George Mason University, Chemistry Department, Fairfax, VA 22030

**Abstract**
Structural relaxation times and viscosities for non-associated liquids and polymers are a unique function of the product of temperature, $T$, times specific volume, $V$, with the latter raised to a constant, $\gamma_\tau$. Similarly, for both neat *o*-terphenyl (OTP) and a mixture the entropy for different $T$ and pressures, $P$, collapse to a single curve when expressed versus $TV^{\gamma_S}$, with the scaling exponent for the entropy essentially equal to the thermodynamic Grüneisen parameter. Since the entropy includes contributions from motions, such as vibrations and secondary relaxations, which do not affect structural relaxation, $\gamma_S < \gamma_\tau$. We show herein that removal of these contributions gives a satisfactory accounting for the magnitude of $\gamma_\tau$. Moreover, the relaxation times of OTP are found to be uniquely defined by the entropy, after subtraction from the latter of a $V$-independent component.

## 1. Introduction

Understanding the origin of the dramatic change in the dynamics of liquids and polymers as the glass transition is approached from the equilibrium state remains a central issue in soft condensed matter, notwithstanding more than 50 years of research. Various interpretations have been proposed, based variously on configurational entropy, energy landscapes, and free volume; however, no consensus has emerged. The problem remains unsolved in part due to the diversity of glass-forming materials and the wealth of experimental data that must be addressed. One significant recent finding is that relaxation times measured under different conditions of $T$ and $P$ superpose to form a single curve when plotted versus $TV^{\gamma_\tau}$ [1,2]; that is

$$\tau = f(TV^{\gamma_\tau}) \qquad (1)$$

where $f$ represent an unknown function. The motivation for eq.(1) derives from simulations by Hansen and coworkers [3,4] on a Lennard-Jones 6-12 fluid showing that the glass transition occurs at a constant value of $T_g V_g^4$. Subsequently from NMR measurements Hollander and Prins [5] found that $T_g V_g^2$ is a constant for polypropylene, while Tölle et al. [6] obtained a similar result for the crossover temperature of mode coupling theory in *o*-terphenyl (OTP) with the exponent equal to 4. These results indicate that a characteristic temperature associated with a given value of the relaxation time occurs at a fixed structure factor. Generalizing this result to any relaxation time [1] leads to eq.(1), whose validity has now been demonstrated for more than 50 liquids, including results from dielectric spectroscopy [1,7,8,9,10,11,12], neutron [13] and light scattering [14,15], viscosity measurements [14,16], and simulations [17,18]. Eq. (1) has also been found to be valid for the normal mode in polymers [19,20] and for the component dynamics in polymer blends [21,22]. A breakdown of the scaling can be expected wherever the material itself changes (chemically) with *T* or *P* – an example being strongly hydrogen bonded materials such as water and glycols [16].

The idea underlying eq.(1) is that for local properties the intermolecular potential can be represented by a two-body, repulsive power law [23,24,25]

$$U(r) \sim r^{-3n} \quad (2)$$

where *r* is the intermolecular separation and *n* a constant related to the steepness of the potential. To the extent eq. (2) is accurate, thermodynamics properties in general will depend only on the scaled variable $TV^n$ [23,24]. This scaling breaks down for the equation of state, due to neglect of the longer range attractions [26]. However, we have shown for three glass-forming materials, propylene carbonate, salol, and polyvinylacetate, that the entropy is well represented by [27]

$$S = f(TV^{\gamma_S}) \quad (3)$$

with the scaling exponent numerically equal to the Grüneisen parameter, $\gamma_G$. This result, $\gamma_S = \gamma_G$, can be derived from eq.(2) [26] or from thermodynamics using

$$\gamma_G = \frac{V\alpha_P}{C_V \kappa_T} \quad (4)$$

where $\alpha_P$ is the isobaric thermal expansivity, $\kappa_T$ the isothermal compressibility, and $C_V$ the isochoric heat capacity. If $\gamma_G$ is independent of *V* (or equivalently $\left.\frac{\partial P}{\partial T}\right|_V \propto \frac{1}{V}$ [28]) it follows that the entropy is

$$S(T,V) = C_V \ln\left(TV^{\gamma_G}\right) + S_{ref} \quad (5)$$

The condition $TV^{\gamma_G} = const$ corresponds to an adiabatic transformation reminiscent of an ideal gas. Interestingly, the scaling exponent for the relaxation times, $\gamma_\tau$, is about threefold larger than $\gamma_S$ [27].

In this paper we test eq.(3) for a prototypical liquid, OTP, and for a mixture of OTP with o-phenyl phenol (OPP). This mixture has a substantially larger value of $\gamma_\tau$ than OTP [29], and as we show herein, the scaling exponent for the entropy is correspondingly larger. We also examine a means to correct the Grüneisen parameter for contributions from motions not involved in structural relaxation, whereby consistency between $\gamma_\tau$ and $\gamma_G$ (and thus $\gamma_S$) is achieved.

## 2. Results

*2.1 OTP*

Quasielastic neutron scattering [13], transverse Brillouin dynamic light scattering [14], and viscosity measurements [14] all indicate a value of $\gamma_\tau = 4.0$ for OTP, consistent with $n = 4$ in eq. (2) [30]. (The reported value for dielectric relaxation times is slightly larger, $\gamma_\tau = 4.25$ [14].) To evaluate the dependence of the entropy on $T$ and $V$, we calculate $S$ using

$$S(T,P) = S_{ref}(T_{ref}, P=0) + \int_{T_{ref}}^{T} \frac{C_P}{T} dT - \int_{0}^{P} \left.\frac{\partial V}{\partial T}\right|_P dP \qquad (6)$$

where $C_p$ is the isobaric heat capacity and $S_{ref}$ the entropy at a reference temperature $T_{ref}$ and $P = 0$. Thus we calculate $S(T,0) - S_{ref}$ at zero pressure using the atmospheric heat capacity data for OTP reported by Chang and Bestul [31]. The results are shown as a function of volume over the range from $T_g \sim 247$ K to $T = 355$ K in the inset to Figure 1. Using the reported equation of state data for OTP [32], we evaluate the third term in eq.(6) for 5 temperatures spanning this range, with the obtained values of $S(T,P) - S_{ref}$ included in the inset.

Next now replot these data as a function of $TV^{\gamma_S}$, with the exponent adjusted to give collapse of the S(T,P) – S$_{ref}$ onto a single master curve. Superpositioning is achieved for $\gamma_S = 1.2$, which is substantially smaller than the value of 4 for $\gamma_\tau$.

We compare this exponent to $\gamma_G$, based on the connection between these quantities [26,28]. Note that while $\gamma_G$ is a strong function of temperature close to 0 K and well below $T_g$, at higher temperature it becomes approximately constant [33,34]. The thermodynamic Grüneisen parameter is defined in eq.(4) [33], with the isochoric heat capacity obtained from $C_p$ using the thermodynamic relation

$$C_V = C_P - \frac{TV\alpha_P^2}{\kappa_T} \tag{7}$$

The result is $\gamma_G = 1.2$ for OTP at $T_g$ and zero pressure, in agreement with the scaling exponent for the entropy.

*2.2 OTP/OPP*

Dielectric relaxation times for a mixture of OTP and OPP (2:1 ratio) were found to scale for a value of the exponent $\gamma_\tau = 6.2$ [29]. Takahara et al. [35] reported heat capacity, thermal expansivity and compressibility measurements at both ambient and P = 28.8 MPa for this same mixture, enabling the entropy to be determined directly for these two pressures. The results are shown in the inset of Figure 2. (Note that both data sets are isobars, whereas Fig. 1 for OTP shows an isobar and 4 isotherms.) When replotted versus $TV^{1.6}$, the data collapse to a single curve. The entropy scaling exponent is smaller by a factor of almost 4 than $\gamma_\tau$. From eq. (4) the Grüneisen parameter for the mixture is calculated to be 1.30 at $T_g$, close to the value of $\gamma_S$ for the mixture.

*2.3 Excess contributions to $\gamma_G$ and S*

The superpositioning of the entropy for the two materials supports identification of the scaling exponent $\gamma_S$ with the Grüneisen parameter; however, for both liquids $\gamma_S$ is 3 – 4 times smaller than the scaling exponent for the relaxation times. This difference is due to the contribution to the entropy from motions not involved in structural relaxation; i.e., vibrations and local secondary processes. This is a problem common to assessments of configurational entropy models of the glass transition [36,37,38,39]. The configurational entropy is usually unavailable, so the total entropy is used after subtraction of the crystal entropy. This subtraction assumes equivalent *T* and *P* dependences for the excess contributions to the total and the crystal entropies, with the expectation that the resulting "excess" entropy is at least proportional to the configurational entropy [40,41].

We have recently proposed a different approach to determine the configurational entropy, $S_c$, from the total entropy, without the need of subtracting the crystal entropy [27]. The small value of the scaling exponent for the entropy connotes a relatively weak volume dependence;

thus, to a first approximation that portion of the entropy not involved in structural relaxation (which we designate $S_0$) can be taken to have a negligible dependence on V; that is, $S_0(T,V) \sim S_0(T)$. This implies that for constant $\tau$, and presumably constant configurational entropy, $S_0 \sim S(T_g)$ (where $T_g$ is the temperature corresponding to a fixed value of $\tau$). For salol and PVAc, it has indeed been shown that their respective $\tau(T,V)$ are essentially a single function of $S(T,V) - S_0(T)$, with the latter defined as described herein [27].

We test this idea for OTP using the dielectric relaxation data reported by Naoki and coworkers [42], who measured $\tau$ over the range $257 < T$ (K) $< 290$ for P from ambient to 79 MPa. Their relaxation times were well described by a generalized Vogel-Fulcher equation having the form [42]

$$\tau = 8.907 \times 10^{-22} \exp\left(\frac{3779 + 3.43P}{T - 170 - 0.19P}\right) \tag{8}$$

Using this equation, the T and P for which $\tau = 1$ were determined and the corresponding values of $S(\tau=1s) - S_{ref}$ ($= S_0 - S_{ref}$) obtained (lower inset to Figure 3); the data exhibits a linear temperature dependence, $S_0 - S_{ref} = -207.1 \pm 0.3 + (0.84 \pm 0.001)T$. The configurational entropy, $S_c = S - S_0$ is then obtained by subtracting from $S - S_{ref}$ the linear fit of $S_0 - S_{ref}$ (solid symbols in the upper inset to Fig. 3). In the main part of Fig. 3, the relaxation times for OTP are plotted versus $S - S_0$. It can be seen that the $\tau$ for different thermodynamic conditions (varying T at constant P and varying P at constant T) fall on a single curve; that is, the relaxation times are uniquely defined by the entropy, after subtraction of the V-independent component of S.

This approach is equivalent to considering that in the liquid state during an isothermal compression or expansion, the entropy change is purely configurational; that is, the unoccupied or "free" volume has to be removed before vibrational or local intramolecular motions are affected. Thus,

$$\left.\frac{\partial S_c}{\partial V}\right|_T = \left.\frac{\partial S^{liq}}{\partial V}\right|_T - \left.\frac{\partial S^{vib}}{\partial V}\right|_T \approx \left.\frac{\partial S^{liq}}{\partial V}\right|_T \tag{9}$$

We therefore calculate $S_c$ as the entropy in excess to the glass, starting from the differential form

$$dS_c = \left(\left.\frac{\partial S^{liq}}{\partial T}\right|_V - \left.\frac{\partial S^{xstal}}{\partial T}\right|_V\right) dT + \left(\left.\frac{\partial S^{liq}}{\partial V}\right|_T\right) dV = \frac{\Delta C_V}{T} dT + \left.\frac{\partial P}{\partial T}\right|_V^{liq} dV \tag{10}$$

where we have used eq.(9) together with one of the Maxwell relations, with $\Delta C_V = C_V^{liq} - C_V^{xstal}$. Defining the Grüneisen parameter for $S_c$, $\gamma_G^{exc}$ as

$$\gamma_G^{exc} = \frac{V\alpha_P^{liq}}{\Delta C_V \kappa_T^{liq}} \tag{11}$$

equation (10) can be rewritten as

$$dS_c = \Delta C_V \left( \frac{dT}{T} + \gamma_G^{exc} \frac{dV}{V} \right) \tag{12}$$

If $\gamma_G^{exc}$ and $\Delta C_V$ are independent respectively of $V$ and of $T$, eq.(12) can be integrated giving similarly to eq.(5)

$$S_c = \Delta C_V \ln\left(TV^{\gamma_G^{exc}}\right) + const \tag{13}$$

Therefore $S_c$ should scale when plotted as a function of the variable $TV^{\gamma_G^{exc}}$. In Table 1 are listed the values of $\gamma_G^{exc}$ calculated according to eq.(11). The crystal data for OTP were taken from refs. 31 and 32. The obtained result is $\gamma_{G,ex} = 3.7$. This value is only 7% smaller than $\gamma_\tau$, whereas there is a factor of 3.3 difference between $\gamma_G$ and $\gamma_\tau$.

We cannot extend this exact procedure to the OTP/OPP mixture because it does not crystallize (indeed, OPP is added in order to suppress crystallization of the OTP). Therefore, instead we calculate $\gamma_{G,ex}$, the excess heat capacity with respect to the glass rather than the crystal. The result is $\gamma_{G,ex} = 6.9$, which is only 10% larger than $\gamma_\tau$.

The requisite data for this calculation are available also for salol. Previously it was found that the Grüneisen parameter defined by eq.(4), while essentially equal to $\gamma_S$, was about threefold smaller than the scaling exponent for the relaxation times of salol [27]. Using crystal data from ref. 43, eq.(11) gives the "excess" Grüneisen parameter, $\gamma_{G,ex} = 6.2$, which compares favorably to the scaling exponent for the relaxation times, $\gamma_\tau = 5.2$ [1].

## 3. Conclusions

We have shown herein that the total entropy of both OTP and its mixture with OPP is a function of $TV^{\gamma_S}$ with $\gamma_S \sim \gamma_G$. The values of $\gamma_S$ for the two liquids differ substantially from the respective scaling exponents for the relaxation times $\gamma_\tau$. This difference is due to the excess contribution to the entropy, $S_0$, from vibrations and secondary processes. For OTP and salol, we carry out a correction to the Grüneisen parameter, using values of the expansivity, compressibility, and heat capacity of the crystalline state, while for OTP/OPP we use values for the glassy state. The obtained $\gamma_{G,ex}$ are in quite good agreement with the reported values for $\gamma_\tau$.

A method is also illustrated to correct the entropy for the excess contribution by assuming that $S_0$ has a negligible dependence on $V$, whereby $S_0(T) \sim S(T_g)$ with the latter obtained from measurements at elevated pressure. It can then be shown that $\tau$ for OTP depends only on $S - S_0$, a result previously found for salol and PVAc [27]. This in turn means that $S - S_0$ conforms to eq.(1) with the same exponent $\gamma_\tau$. These results affirm that the thermodynamic scaling of the relaxation times (eq.(1)) has an entropic origin, supporting the idea that models of the supercooled dynamics of liquids should be based on the configurational entropy.

**Acknowledgement**

This work was supported by the Office of Naval Research.

**References**

[1] Casalini R and Roland CM 2004 *Phys. Rev. E* **69** 062501.

[2] Roland C M, Hensel-Bielowka S, Paluch M and Casalini R 2005 *Rep. Prog. Phys.* **68** 1405.

[3] Bernu B, Hansen J P, Hiwatari Y and Pastore G 1987 *Phys. Rev. A* **36** 4891.

[4] Rous J N, Barrat J L and Hansen J P 1989 *J. Phys. Cond. Matt.* **1** 7171.

[5] Hollander A G S and Prins K O 2001 J. Non-Cryst. Solids **286** 1.

[6] Tölle A, Schober H, Wuttke J, Randl O G, Fujara F 1998 *Phys. Rev. Lett.* **80** 2374.

[7] Casalini R and Roland CM 2004 *Colloid Polym. Sci.* **283** 107.

[8] Roland C M and Casalini R 2005 *J. Non-Cryst. Solids* **251** 2581.

[9] Alba-Simionesco C, Cailliaux A, Alegria A and Tarjus G. 2004 *Europhys. Lett.* **68** 58.

[10] Urban S and Würflinger A 2005 *Phys. Rev. E.* **72** 021707.

[11] Reiser A, Kasper G and Hunklinger S 2005 *Phys. Rev. B* **72** 094204.

[12] Win K Z and Menon N 2006 *Phys. Rev. E* **73** 040501.

[13] Tölle A 2001 *Rep. Prog. Phys.* **64** 1473.

[14] Dreyfus C, Aouadi A, Gapinski J, Matos-Lopes M, Steffen W, Patkowski A and Pick R M 2003 *Phys. Rev. B* **68** 011204.

[15] Dreyfus C, Le Grand A, Gapinski J, Steffen W and Patkowski A 2004 *Eur. J. Phys.* **42** 309.

[16] Roland C M, Bair S and Casalini R 2006 *Macromolecules*, in press; cond-mat/0607611.

[17] Tsolou G, Harmandaris V A, and Mavrantzas V G 2006 *J. Chem. Phys.* **124** 084906.

[18] Budzien J, McCoy J D and Adolf DB 2004 *J. Chem. Phys.* **121** 10291.

[19] Roland C M, Casalini R and Paluch M 2004 *J. Polym. Sci. Polym. Phys. Ed.* **42** 4313.

[20] Casalini R and Roland C M 2005 *Macromolecules* **38** 1779.

[21] Roland C M and Casalini R 2005 *Macromolecules* **38** 8729.

[22] Roland, K.J. McGrath, and R. Casalini, Macromolecules **39** 3581 (2006).

[23] March N H and Tosi M P 2002 *Introduction to the Liquid State* (Singapore: World Scientific).

[24] Hoover W G and Ross M 1971 *Contemp. Phys.* **12** 339.

[25] Chandler D, Weeks J D and Andersen H C 1983 *Science* **220** 787.

[26] Roland C M, Feldman J L and Casalini R 2006 *J. Non-Cryst. Sol.*, in press; cond-mat/0602132.

[27] Casalini R and Roland C M 2006 *Phil. Mag.* in press.

[28] Casalini R, Mohanty U and Roland C M 2006 *J. Chem. Phys.* **125** 014505.

[29] Roland C M , Capaccioli S, Lucchesi M and Casalini R 2004 *J. Chem. Phys.* **120** 10640.

[30] Lewis L J and Wahnström G 1994 *Phys. Rev. E* **50** 3865.

[31] Chang S S and Bestul A B 1972 *J. Chem. Phys.* **56** 503.

[32] Naoki M and Koeda S 1989 *J. Phys. Chem.* **93** 948.

[33] Hartwig G 1994 *Polymer Properties at Room and Cryogenic Temperatures* (New York: Plenum Press), chapter 4.

[34] Curro J G 1973 *J. Chem. Phys.* **58** 374.

[35] Takahara S, Ishikawa M, Yamamuro O and Matsuo T 1999 *J. Phys. Chem. B* **103** 792; *ibid.* 3288.

[36] Johari G P 2003 *J. Chem. Phys.* **119** 635.

[37] Richert R and Angell C A 1998 *J. Chem. Phys.* **108** 9016.

[38] Yamamuro O, Tsukushi I, Lindqvist A, Takahara S, Ishikawa M and Matsuo T 1998 J. Phys. Chem. B **102** 1605.

[39] Prevosto D, Lucchesi M, Capaccioli S, Casalini R and Rolla P A 2003 *Phys. Rev. B* **67** 174202.

[40] Angell C A  Borick S 2002 *J. Non-Cryst. Sol.* **307** 393.

[41] Prevosto D, Lucchesi M, Capaccioli S, Casalini R and Rolla P A 2003 *Phys. Rev. B* **67** 174202.

[42] Naoki M, Endou H and Matsumoto K 1987 *J. Phys. Chem.* **91** 4169.

[43] Comez L, Corezzi S, Fioretto D, Kriegs H, Best A and Steffan W 2004 *Phys. Rev. E* **70** 011504.

Table 1. Scaling exponents and Grüneisen parameters

|          | $\gamma_\tau$ | $\gamma_S$ | $\gamma_G$ | $\gamma_{G,ex}$ |
|----------|------|------|------|------|
| OTP      | 4.0 [a] | 1.2 | 1.2 | 3.7 |
| OTP/OPP  | 6.2 [b] | 1.6 | 1.3 | 6.9[c] |
| salol    | 5.2 [d] | 1.7 [e] | 1.9 [f] | 6.2 |

[a] ref. 13 and 14  
[b] ref. 29  
[c] glass values used since there is no crystalline state  
[d] ref. 1  
[e] ref. 27  
[f] ref. 28

**Figure Captions**

Figure 1. Scaled plot of the total entropy minus the entropy at the ambient pressure $T_g$ for OTP (5 isotherms and 1 isobar). The inset shows the data versus specific volume. The entropies were calculated at P=0.1 MPa from the heat capacity [31] and at higher pressures from the equation of state [32] using eq.(6).

Figure 2. Scaled plot of the total entropy minus the entropy at the ambient pressure $T_g$ for a 2:1 mixture of OTP and OPP (2 isobars). The inset shows the data versus specific volume. The entropies were calculated from the heat capacity at the two pressures as reported in ref. [35].

Figure 3. OTP relaxation times [42] versus the total entropy minus the contribution from motions not involved in structural relaxation (symbols as in Fig. 1). This contribution, $S_0$, was taken to equal the value of the entropy at $T_g(P)$ (with $S_0(T)$ assumed volume independent). The top inset compares $S - S_0$ and $S - S_{ref}$ (with $S_{ref}$ a constant equal to $S(T_g, 0.1$ MPa$)$). The lower inset shows $S_0 - S_{ref}$ fitted to the linear equation given in the text.

Figure 1.

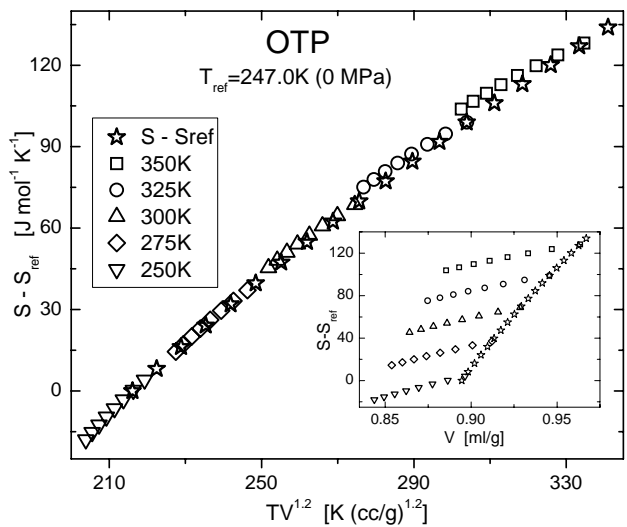

Figure 2.

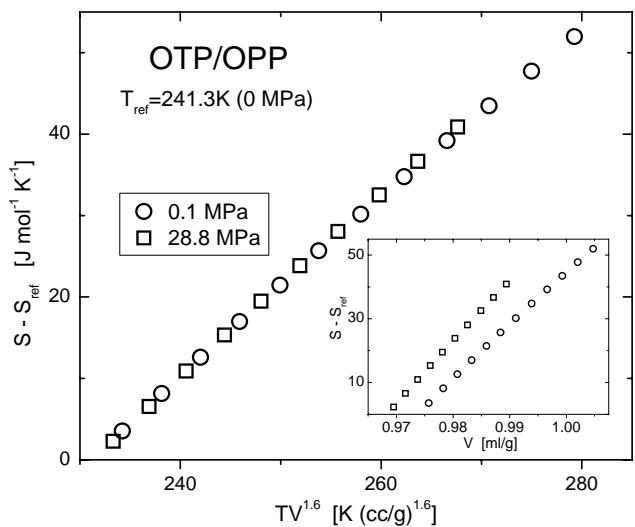

Figure 3

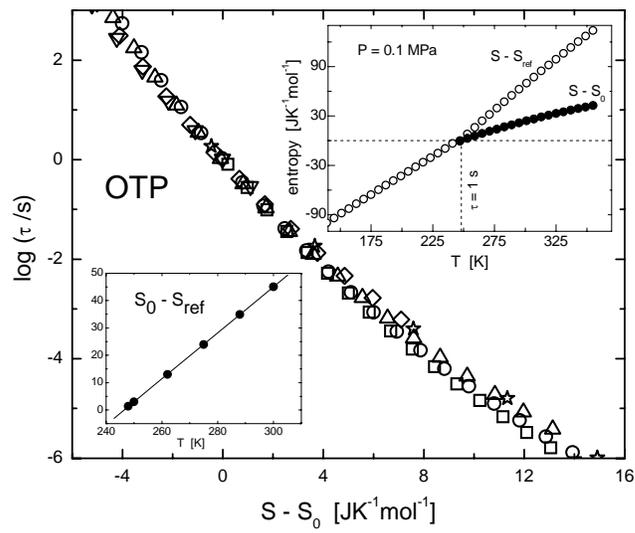